\documentclass[pre,a4paper,onecolumn,nofootinbib]{revtex4}
\usepackage{amsmath}
\usepackage{bbm}
\usepackage{psfrag}
\usepackage{latexsym}
\usepackage{graphicx}
\usepackage{subfigure}% Include figure files
\usepackage{amsfonts}
\newcommand{\be}{\begin{equation}}
\newcommand{\ee}{\end{equation}}

\begin{document}

\title{New spectral relations  between products and powers of  isotropic random matrices}

\author{Z. Burda$^{1,2}$}\email{zdzislaw.burda@uj.edu.pl}
\author{M.A. Nowak$^{1,2}$}\email{nowak@th.if.uj.edu.pl}
\author{A. Swiech$^{1}$}\email{artur.swiech@uj.edu.pl}

\affiliation{$^{1}$Marian Smoluchowski Institute of Physics and 
$^{2}$Mark Kac Complex Systems Research Center \\
Jagiellonian University, Reymonta 4, 30-059 Krak\'ow, Poland}

\date{\today}

\begin{abstract}
We show that the limiting eigenvalue density of the product of $n$ identically distributed random matrices from an isotropic unitary ensemble (IUE) 
is equal to the eigenvalue density of $n$-th power of a single matrix
from this ensemble,  in the limit when the size of the matrix tends to infinity.
Using this observation one can derive the limiting density of the product of $n$ independent identically distributed non-hermitian matrices with
unitary invariant measures. In this paper we discuss two examples: 
the product of $n$ Girko-Ginibre matrices and the product of $n$ truncated unitary matrices. We also provide an evidence that the result holds also for 
isotropic orthogonal ensembles (IOE).

\medskip


\end{abstract}

\maketitle

\section*{Introduction}
Free probability theory is a fusion of non-commutative probability theory and the concept of free independence. Since 1991, when the link between the free probability theory and random matrix theory was established~\cite{vdn}, several new results have been proven in an easy and powerful way in the limit of infinitely large random matrices~\cite{hp,agz,ns}. 
In this note we demonstrate a simple albeit quite counterintuitive result that the spectral density of the product of $n$ free, identically distributed 
random matrices from an isotropic unitary ensemble (IUE) is equal to the spectral density of the $n$-th power of a single matrix from this ensemble in the limit of infinite matrix size. The proof is based on the multiplicative properties of the S-transform and the Haagerup-Larsen theorem~\cite{hl}. 

The motivation for the present work comes from the observation made
in \cite{bjw,agt1,agt2} that the eigenvalue density of $n$ independent 
Girko-Ginibre~\cite{g1,g2} 
matrices is identical to the eigenvalue density of the $n$-th power of a single Girko-Ginibre matrix in the limit of infinite size. 
This observation leads to the question whether this is a feature 
of only this particular class of matrices or if there exists a larger class of 
matrices that have this property. In the present paper we show that there is indeed 
a larger class of matrices sharing this property - a class of random isotropic matrices.
We begin with defining isotropic matrices. Then we 
present the main result in detail and its derivation. Finally 
we outline a few sample applications, related to the recent interest in the 
literature. In particular we apply our result to the product of Ginibre-Girko 
matrices and rederive the density known from \cite{bjw,agt1,agt2}. We also consider
classes of truncated unitary and orthogonal matrices and compare 
our predictions to Monte Carlo simulations, which allow us to identify finite size corrections. We conclude the paper with a short summary. 

\section*{Isotropic random matrices}

It is convenient to introduce the concept of isotropic random matrices in
analogy to isotropic complex random variables $z$ that have 
a circularly symmetric probability distribution depending only on 
the module $|z|$.  Using polar decomposition, one can write 
$z=r e^{i\phi}$ where  $r$ is a real non-negative random variable 
and $\phi$ is a random variable (phase) with a uniform distribution on $[0,2\pi)$. Isotropic random matrices are defined by a straightforward generalization of isotropic complex random variables. A square $N\times N$ matrix $x$ is said to be isotropic random matrix if it has a polar decomposition $x=hu$ in which $h$ is a positive semi-definite Hermitian random matrix and $u$ is a unitary random matrix independent of $h$ and distributed on the unitary
group $U(N)$ with the Haar measure. In short, $u$ is a 
Haar unitary matrix. The random matrix $h$ plays the role of the 
radial part of $x$. Such random matrices form an ensemble of isotropic unitary matices (IUE). An example is an ensemble generated by the partition function \cite{fz,fsz}:
\be
Z = \int Dx \ e^{-N {\rm Tr} V(x^\dagger x)}
\ee
where $Dx = \prod_{ij} d ({\rm Re} x_{ij}) d({\rm Im} x_{ij})$ is a flat measure, and $V(a)$ is a polynomial in $a$. Another natural class of IUE
matrices are matrices of the form 
$x = v d u$ where $d$ is an $N \times N$
diagonal matrix having real positive random eigenvalues with the given
probability distribution and $v$ and $u$ are two independent Haar unitary matrices
on the unitary group $U(N)$. By analogy one can also consider isotropic orthogonal 
ensemble (IOE) given by the decomposition $x=so$ with $s$ being a positive 
semidefinite real symmetric matrix and $o$ being a Haar orthogonal matrix. 
In this case, when one considers an ensemble given by a partition function 
like that given above, one has to replace $x$ by a real matrix. 

In mathematical literature, isotropic matrices for $N\rightarrow \infty$ are called  $R$-diagonal \cite{ns2}. In this note we prefer to call them
isotropic (or IUE, IOE) in the large $N$ limit. IUE matrices have an eigenvalue distribution independent of the polar angle on the complex plane. 
In the limit when  the matrix size $N\rightarrow \infty$ one can find an explicit relation between the eigenvalue density of the matrix 
$h^2$ and of the matrix $x$ \cite{hl,fz,fsz,gkz}.
We briefly recall this relation below. Let us mention that the angular independence of the eigenvalue density does not imply that the
matrix is isotropic. For example a block diagonal matrix
of the form:
\be
x = \left( \begin{array}{cc} h_1 u_1 & 0 \\ 0 & h_2 u_2 \end{array} \right) 
\ee
where $h_1,h_2$ are independent Hermitian matrices of dimensions $N_1$ and $N_2$,
$N_1+N_2=N$, and $u_1,u_2$ are Haar unitary matrices on $U(N_1)$ an
$U(N_2)$ respectively, has a circularly symmetric eigenvalue density 
in the complex plane, but it is not isotropic. Intuitively, this is because the split into $u_1$ and $u_2$ breaks the isotropy in the whole $U(N)$ group.

\section*{Main result}
The main result of this paper is as follows:  
consider $n$ identically distributed isotropic matrices 
$x_1$, $x_2$, $\ldots$, $x_n$ generated independently from 
a given IUE (isotropic unitary ensemble). In the limit $N\rightarrow \infty$ the eigenvalue density of the product $X_n = x_1 x_2 \ldots x_n$ becomes identical 
as the eigenvalue density of the $n$-th power $x^n$ of a single matrix 
$x$ from this ensemble (e.g. $x=x_1$). In other words, 
the probability that a randomly chosen eigenvalue of $X_n$ lies within a circle of radius $r$: ${\rm Prob}(\lambda_{X_n} < r)$ approaches for $N\rightarrow \infty$ the probability that a randomly chosen eigenvalue of $x^n$ lies within the same circle: ${\rm Prob}(\lambda_x^n<r)$.
One can use this observation to derive the eigenvalue density of the product $X_n = x_1 x_2 \ldots x_n$ if the eigenvalue density of $x$ is known. In particular one can immediately show that the eigenvalue distribution of the product of $n$ independent Girko-Ginibre matrices has a simple form:
\be
\label{ggn}
\rho(z,\bar{z}) = \frac{1}{\pi n} |z|^{-2+2/n} \qquad \mbox{for} \quad |z|\le 1 
\ee
and zero for $|z|>1$, in agreement with \cite{bjw,agt1,agt2,bjn,b1,b2,os}.
It is interesting to note that the matrices $X_n X_n^\dagger$ obtained from
the products $X_n$ of Girko-Ginibre matrices generate a Fuss-Catalan
family of distributions \cite{cnz} that have however a much more complicated
limiting eigenvalue density \cite{pz}. Another interesting case is
the product of $n$ independent truncated unitary matrices \cite{sz} that is
\be
\label{tun}
\rho(z,\bar{z}) = \frac{\kappa}{n\pi} |z|^{-2+2/n}(1-|z|^{2/n})^{-2}
\qquad \mbox{for} \quad |z|\le \left(\frac{1}{1+\kappa}\right)^{n/2}
\ee
and zero otherwise. The truncated matrices
have dimensions $N\times N$. They are obtained by removing $L$
columns and $L$ rows from the original $(N+L) \times (N+L)$ 
Haar unitary random matrix. The result holds for 
$N\rightarrow\infty$ and $\kappa=L/N$ fixed.

This is a counterintuitive result, so let us stress that it only holds
in the limit $N\rightarrow \infty$. For finite $N$ the eigenvalue 
distributions of the product of $x_1 \ldots x_n$ and of the power $x^n$ differ.
The difference however disappears when $N$ tends to infinity, as we illustrate it below.

\section*{Derivation}

Consider an IUE ensemble of random matrices $x=hu$ of dimensions $N \times N$.
In the large $N$ limit the random matrices can be represented as free random variables and one can use the Haagerup-Larsen theorem \cite{hl} that relates 
the eigenvalue density of $x$ to the eigenvalue density of $h^2$ by 
the following formula:
\be
\label{HL1}
S_{h^2}\left(F_x(r) - 1\right) = \frac{1}{r^2} \ .
\ee
where $F_x(r)$ is the cumulative density function for the density
of eigenvalues of $x$ on the complex plane 
and $S_{h^2}(x)$ is the S-transform for the matrix $h^2$.
The cumulative density function
\be
\label{Fx}
F_x(r) = \int_{|z|\le r} d^2 z \rho_x(z,\bar{z}) = 
2\pi \int_0^r d s s \varrho_x(s) 
= \int_0^r d s p_x(s),
\ee
can be interpreted as the fraction of eigenvalues of $x$ in the circle 
of radius $r$ centered at the origin of the complex plane. It is
related to the eigenvalue density $\rho_x(z,\bar{z})=\varrho_x(|z|)$
that depends on the distance from the origin $|z|$. 
The integrand $ds p_x(s) = 2\pi ds s \varrho_x(s)$ 
is interpreted as the probability of finding eigenvalues of $x$ 
in a narrow ring of radii $|z|$ and $|z| + d|z|$:
\be
\label{density}
F'_x(r) =  p_x(r) = 2\pi r \varrho_x(r) \ .
\ee
The prime denotes the derivation with respect to the radial variable.  The cumulative density function $F_x(r)$ enters equation (\ref{HL1})
as an argument of the S-transform $S_{h^2}(z)$ 
that is related to the eigenvalue density $\rho_{h^2}(\lambda)$
of the matrix $h^2$ (see Appendix A). The Haagerup-Larsen theorem states 
also \cite{hl,fz,fsz,gkz} that the support of the eigenvalue
density of $x$ is a ring of radii $R_{\rm min}$ and $R_{\rm max}$ or a disk (if
$R_{\rm min}=0$):
\be
R^2_{\rm max} = \int_0^\infty d\lambda \lambda \rho_{h^2}(\lambda) \quad , \quad
R^{-2}_{\rm min} = \int_0^\infty d \lambda \lambda^{-1} \rho_{h^2}(\lambda) \ .
\ee
Let us make few comments. For an $R$-diagonal (isotropic) matrix $x$ given by 
the radial decomposition $x=hu$, where $h$ is Hermitian
and $u$ is a Haar unitary matrix, the two matrices
$xx^\dagger = h^2$ and $x^\dagger x = u^\dagger h^2 u$ 
have identical eigenvalues and therefore 
the S-transforms for $xx^\dagger$ and $x^\dagger x$ are identical:
$S_{xx^\dagger}(z)=S_{x^\dagger x}(z) = S_{h^2}(z)$.
This means that (\ref{HL1}) can be written as
\be
\label{HL2}
S_{x^{\dagger}x}\left(F_x(r) - 1\right) = \frac{1}{r^2} \ .
\ee
Let us now apply this equation to the product of $n$ identically
distributed $R$-diagonal (isotropic)  matrices $X_n = x_1\ldots x_n$.
The resulting matrix has an identical eigenvalues as 
$ H_n u_n$, where $H^2_n = X_n^\dagger X_n$
so we can apply (\ref{HL2}) replacing in this equation
$x$ by $X_n$:
\be
\label{HL3}
S_{X_n^{\dagger}X_n}\left(F_{X_n}(r) - 1\right) = \frac{1}{r^2} \ .
\ee
The S-transform for the matrix $X_n^{\dagger}X_n$ which appears
in the last equation can be substituted by the S-transforms for
individual terms in the product. Indeed, writing
\be
X_n^{\dagger}X_n = x^\dagger_n X_{n-1}^\dagger X_{n-1} x_n
\ee
where $X_{n-1}=x_1 \ldots x_{n-1}$ we see that
\be
S_{X_n^{\dagger}X_n} = S_{X_{n-1}^\dagger X_{n-1}} S_{x_n^\dagger x_n}
\ee
since due to the cyclic properties of trace the moments of
$x^\dagger_n X_{n-1}^\dagger X_{n-1} x_n$ are identical as those
of $x_n x^\dagger_n X_{n-1}^\dagger X_{n-1}$ and moments of
$x_n x^\dagger_n$ as those of $x^\dagger_n x_n$. Applying the last equation
recursively we eventually obtain
\be
S_{X_n^{\dagger}X_n} = \prod_{i=1}^n S_{x_i^\dagger x_i} \ .
\ee
Taking into account that all $x_i$ are identically distributed
and having the same S-transform (that we denote by $S_{x^\dagger x}$)
we can write the last equation as
\be
S_{X_n^\dagger X_n} = S^n_{x^\dagger x} \ .
\ee
Inserting this into (\ref{HL3}) we have
\be
S_{x^\dagger x}\left(F_{X_n}(r) -1\right) = \frac{1}{r^{2/n}} \ .
\ee
This equation has an identical form as (\ref{HL2}) except that on the
left hand side $F_x(r)$ is replaced by $F_{X_n}(r)$ and on the
right hand side $r$ is replaced by $r^{1/n}$. From this observation
it immediately follows that 
\be
\label{main}
F_{X_n}(r) =  F_x(r^{1/n}) = F_{x^n}(r) \ .
\ee 
The last equality follows from the fact that eigenvalues of
the matrix $x^n$ are equal to the $n$-th power of the corresponding
eigenvalues of $x$: 
$F_{x^n}(r) \equiv 
\mbox{Prob}(|\lambda|^n \le r) = \mbox{Prob}(|\lambda|\le r^{1/n}) \equiv
F_x(r^{1/n})$. So we see that indeed the
product of $n$ identically distributed isotropic 
matrices $X_n = x_1 x_2 \ldots x_n$ has the same eigenvalue distribution
as the $n$-th power $x^n$ of a single matrix in the product. 
In practice, the eigenvalue distribution of $X_n$ can be calculated 
directly from the eigenvalue distribution
of a single matrix $x$ by substituting $r \rightarrow r^{1/n}$
in the cumulative distribution function $F_x(r)$ (\ref{Fx}).
The corresponding eigenvalue densities may be found using 
(\ref{density}). They read
\be
p_{X_n}(r) = \frac{1}{n} r^{1/n-1} p_x(r^{1/n}) 
\ee
and
\be
\varrho_{X_n}(r) = \frac{1}{n} r^{2/n-2} \varrho_{x}(r^{1/n}) \ .
\ee

\section*{Applications}

Let us apply these formulas to a couple of examples.
First consider Girko-Ginibre matrices~\cite{g1,g2} that
have a uniform distribution $\varrho_x(r) = 1/\pi$
inside the unit circle $|z| \le 1$. We have
\be
F_x(r) = 2\int_0^r r' dr' = r^2  \qquad \mbox{for} \quad r\le 1
\ee
and $1$ otherwise. For the product of 
$n$-independent Girko-Ginibre matrices we have (\ref{main})
\be
F_{X_{n}}(r) = r^{2/n} \qquad \mbox{for} \quad r\le 1  
\ee
and one otherwise. Taking the derivative with respect
to $r$ (\ref{density}) we find the corresponding densities:
\be
p_{X_{n}}(r)=\frac{2}{n} r^{2/n-1} \theta(1-r)
\ee
and
\be
\varrho_{X_{n}}(r)=\frac{1}{\pi n} r^{2/n-2} \theta(1-r)
\ee
where $\theta$ denotes the Heaviside step function. 
This result agrees with that obtained using different methods
in \cite{bjw,bjn,b1,b2,os} as mentioned in the introduction  of the paper.

As the second example, we consider the product of $n$ truncated 
unitary matrices \cite{sz}. The cumulative eigenvalue distribution 
of a single matrix from this ensemble is
\be
F_x(r) = \frac{\kappa r^2}{1-r^2} \qquad  \mbox{for} \quad  r\le 
\left(\frac{1}{1+\kappa}\right)^{1/2}
\label{single_trunc}
\ee
and $1$ otherwise. The coefficient $\kappa=L/N$ is the ratio of the number
of rows and columns $L$ removed from a Haar unitary matrix of 
dimensions $(N+L)\times (N+L)$. This truncation leaves a matrix
of dimensions $N\times N$. In Appendix B we show how to derive this
result using free random variables. The corresponding density reads:
\be
\varrho_x(r) = \frac{\kappa}{\pi} (1-r^2)^{-2} \theta\left(\left(\frac{1}{1+\kappa}\right)^{1/2}-r\right) \ .
\ee
Using (\ref{main}) we find the distribution of eigenvalues
for the product of $n$ such matrices:
\be
F_{X_n}(r) = \kappa \frac{r^{2/n}}{1-r^{2/n}} 
\qquad \mbox{for} \quad r\le \left(\frac{1}{1+\kappa}\right)^{n/2}
\ee
and $1$ otherwise. The corresponding eigenvalue density is
\be
\varrho_{X_n}(r) = \frac{\kappa}{n\pi} r^{2/n-2}(1-r^{2/n})^{-2} \theta\left(\left(\frac{1}{1+\kappa}\right)^{n/2}-r\right) \ .
\ee

\section*{Numerical comparison and finite size effects}

\begin{figure}
\centering
\includegraphics[width=59.1mm]{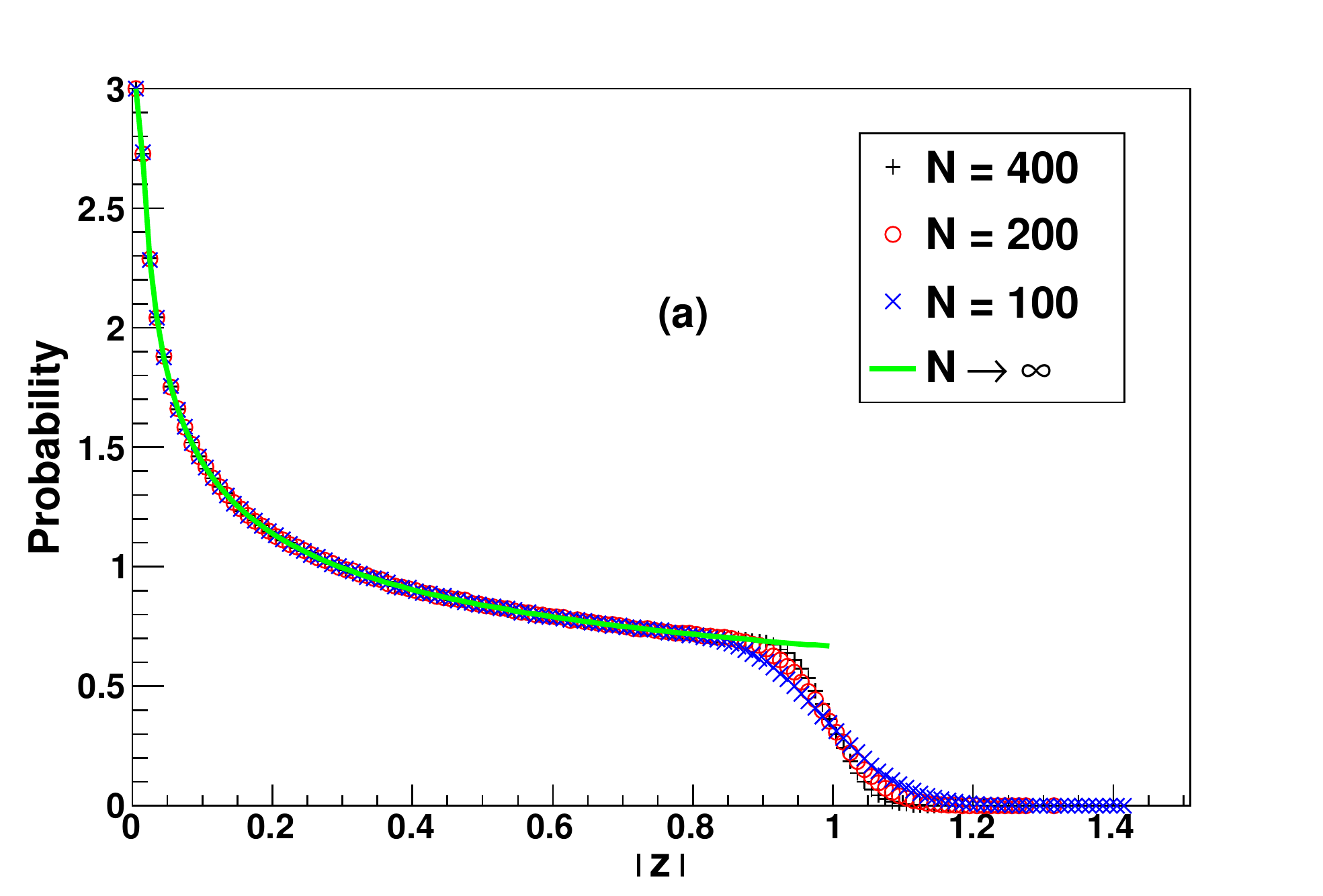} 
\includegraphics[width=59.1mm]{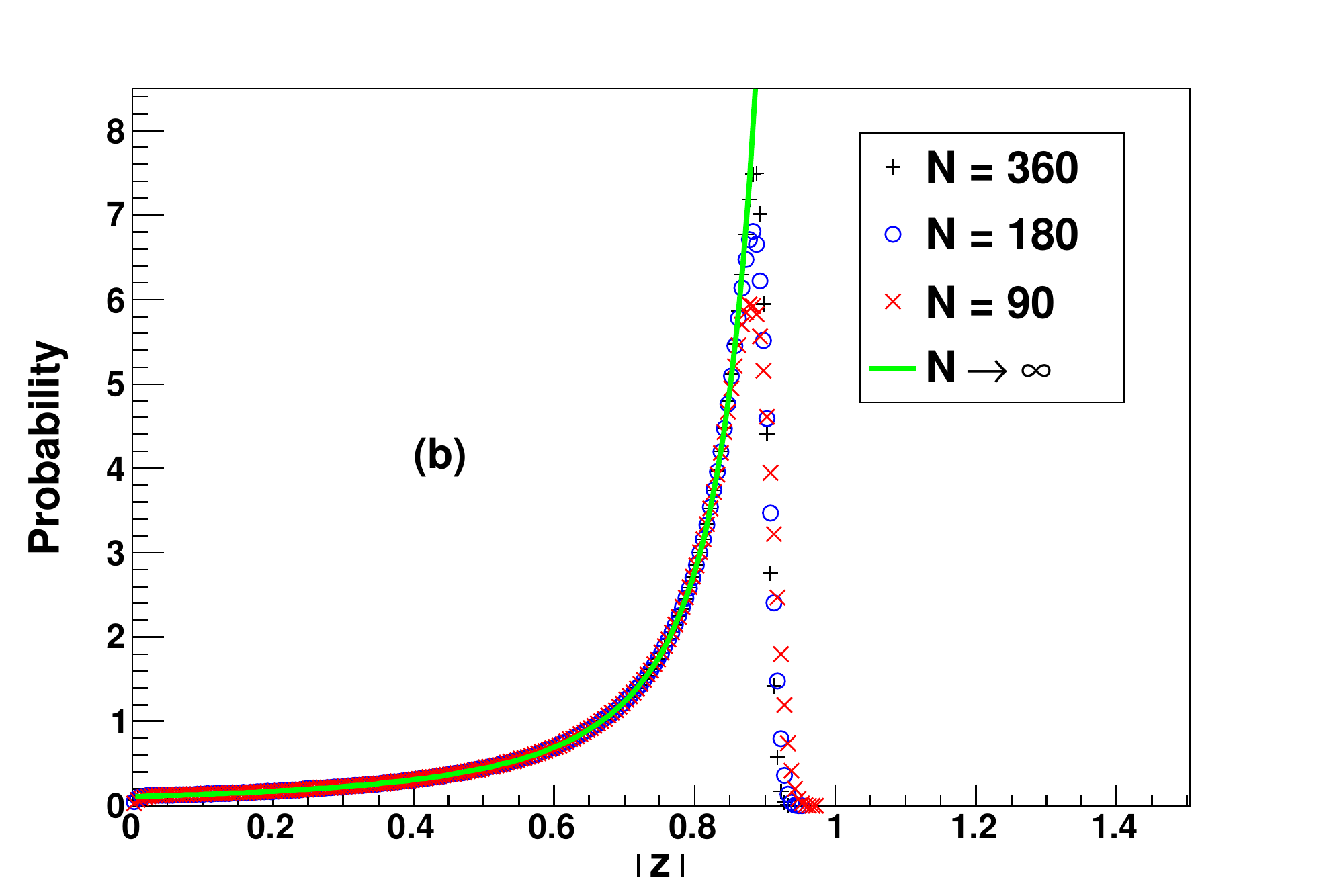} 
\includegraphics[width=59.1mm]{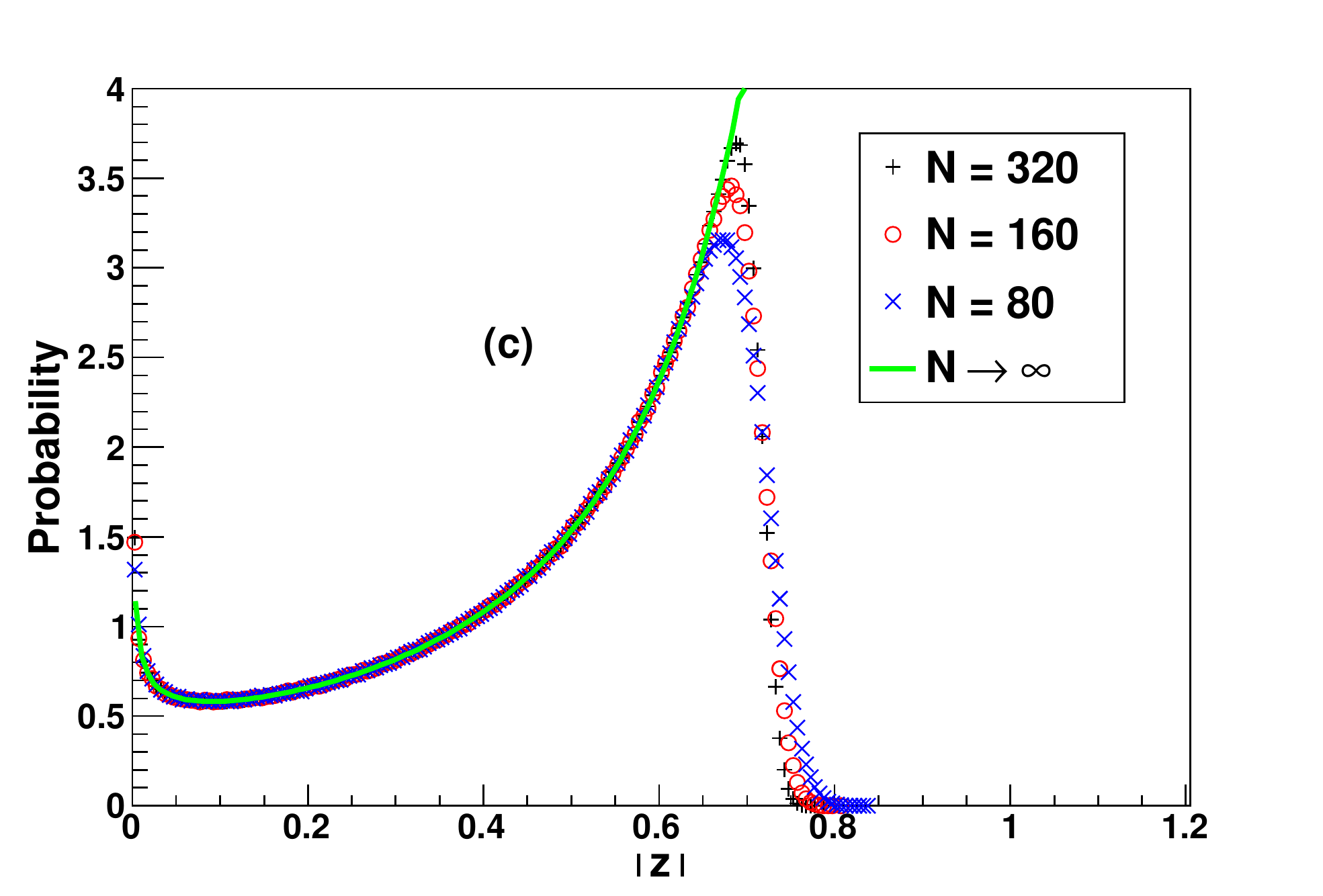}
\caption{(Color online) Numerical verification of theoretical formulas (\ref{ggn}) (a) and (\ref{tun}) (b),(c) for the radial part $p(r)=F'(r)$ (\ref{density}) of the mean spectral density $\rho\left(z,\bar{z}\right)$ of the product of independent matrices. (a) Numerical histograms for the product of 3 independent Gaussian random matrices $N=400$ (black crosses), $N=200$ (red circles) and $N=100$ (blue rotated crosses) compared to  theoretical prediction for $N\rightarrow\infty$ (solid green line). Each histogram is made for $10^{7}$ eigenvalues. The numerical histograms approach theoretical curve as the size of matrices increases. (b) An analogous plot to (a)  for the product of 2 independent truncated unitary matrices with ratio $\kappa=\frac{1}{9}$ and $N=360$ (black crosses), $N=180$ (red circles) and $N=90$ (blue rotated crosses). Each histogram is made for $9\times10^{6}$ eigenvalues. (c) An analogous plot to (a) and (b)  for the product of 3 independent truncated unitary matrices with ratio $\kappa=\frac{1}{4}$ and $N=320$ (black crosses), $N=160$ (red circles) and $N=80$ (blue rotated 
crosses). Each histogram represents  $8\times10^{6}$ eigenvalues.}
\label{fig:numsize}
\end{figure}

\begin{figure}
\centering
\includegraphics[width=59.1mm]{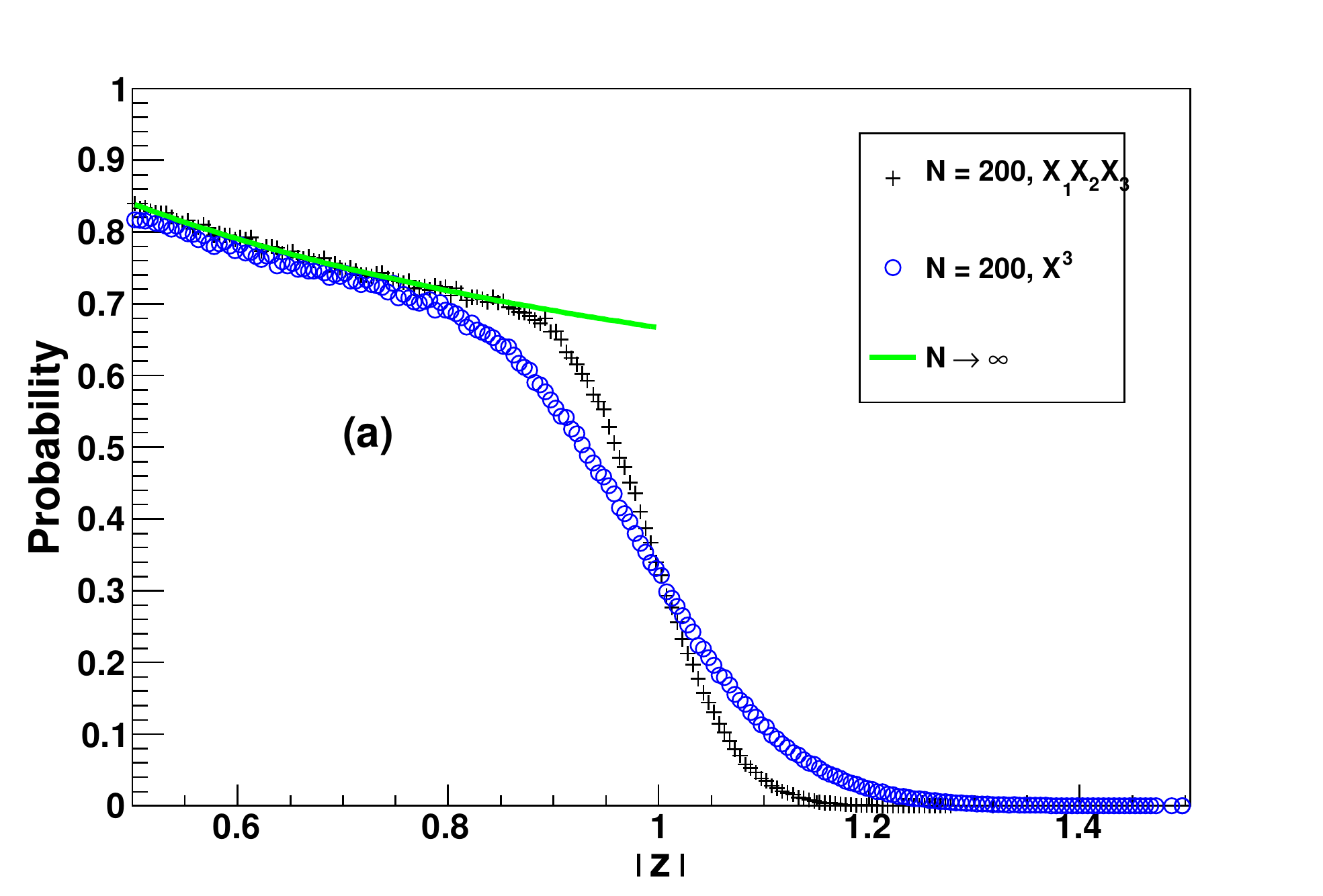} 
\includegraphics[width=59.1mm]{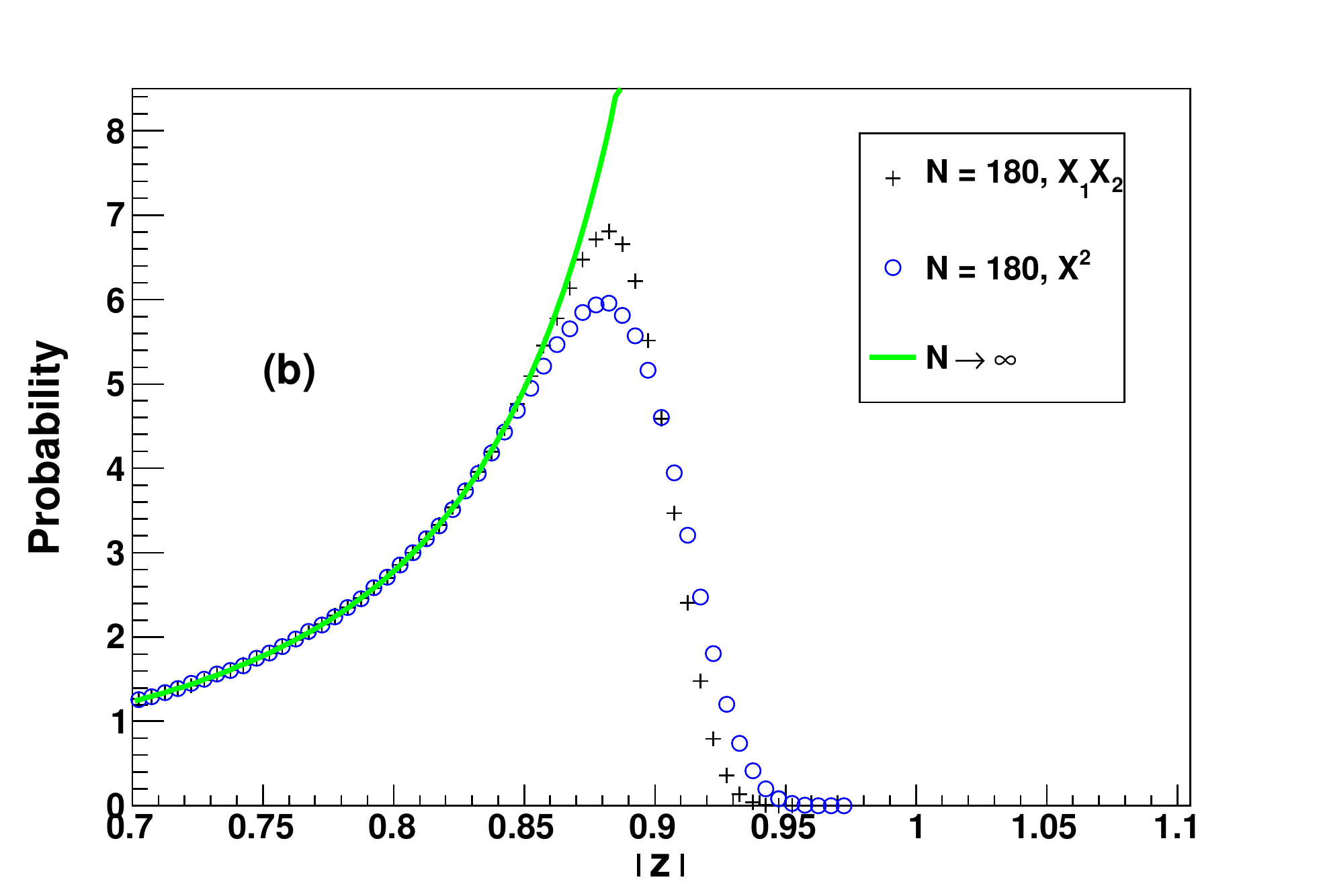} 
\includegraphics[width=59.1mm]{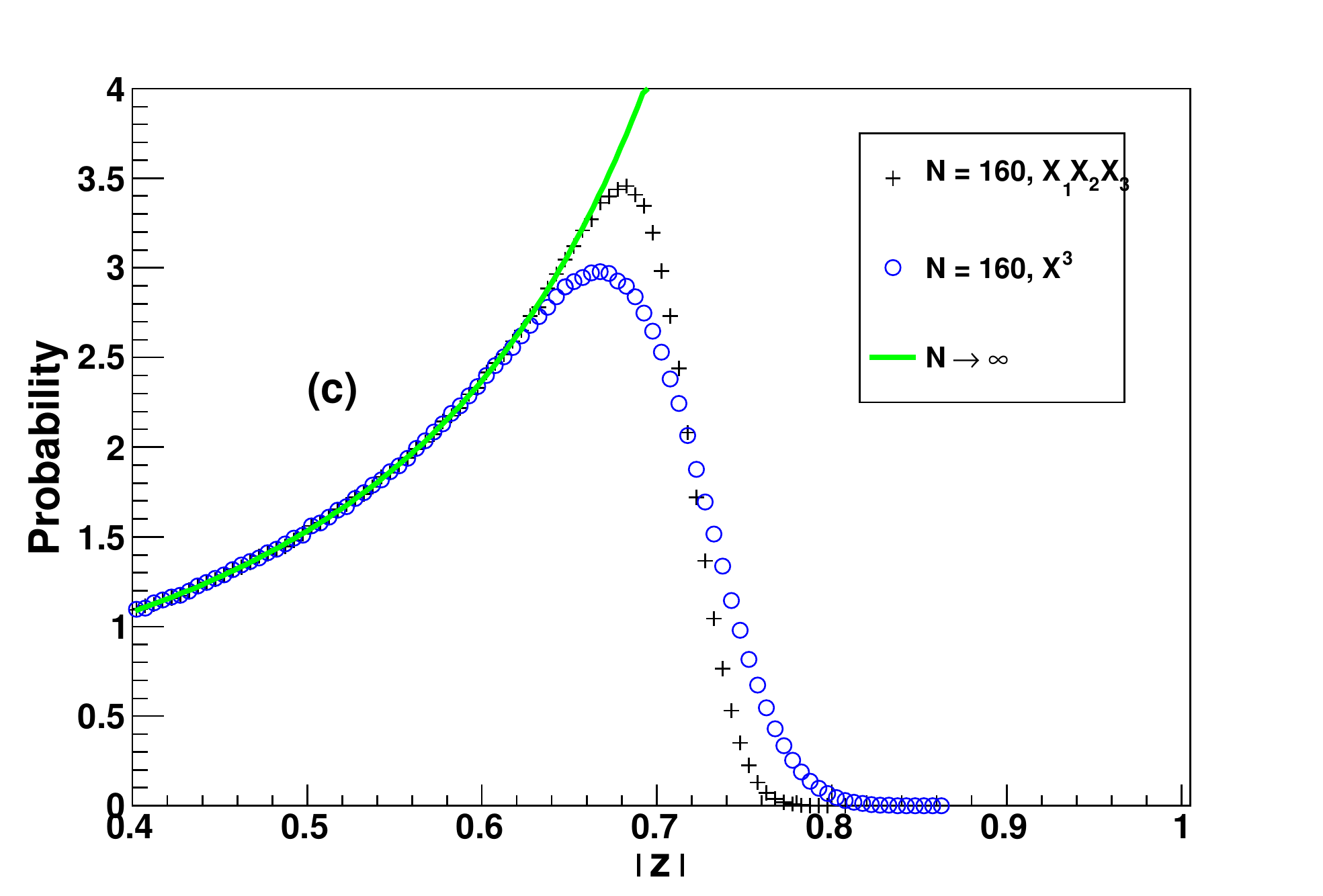}
\caption{(Color online) Numerical analysis of the finite size effects for the radial part $p(r)=F'(r)$ (\ref{density}) of the mean spectral density $\rho\left(z,\bar{z}\right)$ of the product of independent matrices in comparison to the power of a single matrix. (a) Numerical histograms for product of 3 independent Gaussian random matrices $N=200$ (black crosses) and one matrix raised to 3'rd power for $N=200$ (blue circles) compared to theoretical prediction for $N\rightarrow\infty$ (solid green line). Each histogram is made for $10^{7}$ eigenvalues. Plots are zoomed in the region, where the difference in the shape is visible. (b) An analogous plot to (a)  for the product of 2 independent truncated unitary matrices (black crosses) and 2'nd power of a single truncated unitary matrix (blue circles) with ratio $\kappa=\frac{1}{9}$ and $N=180$. Each histogram is made for $9\times10^{6}$ eigenvalues. (c) An analogous plot to (a) and (b) but  the product of 3 independent truncated unitary matrices (black crosses) and 3'rd power of a single truncated unitary matrix (blue circles) with ratio $\kappa=\frac{1}{4}$ and $N=160$. Each histogram is made for $8\times10^{6}$ eigenvalues.}
\label{fig:numprod}
\end{figure}

In order to crosscheck our results, we use Monte-Carlo simulations for generating (sampling) finite size random matrices from ensembles in question. 
An agreement between the analytical formulas (\ref{ggn}) or (\ref{tun}) and numerical results is observed taking into account finite size corrections. The shape of obtained distributions
$p(r) =F'(r)$ (\ref{density}) is shown in the figure~\ref{fig:numsize}. In the $N\rightarrow\infty$ limit, distributions have got compact support, and the sharp drop at the edge is present. For finite $N$ the spectra do not have a sharp threshold -- instead they tend  to zero continuously in an extended crossover region, and the difference between the product of independent matrices and the corresponding
power of a single one is visible in this region (figure~\ref{fig:numprod}). The
eigenvalue density of the product of independent matrices approaches the theoretical curve faster than of the corresponding power of a single matrix.  Only radial distributions $p(r)=F'(r)$ (\ref{density}) are shown, since  eigenvalue densities are circularly symmetric on the complex plane. The shape of the finite size corrections 
for Girko-Ginibre distribution was discussed
in \cite{b1,b2}.

\begin{figure}
\centering
\includegraphics[width=59.1mm]{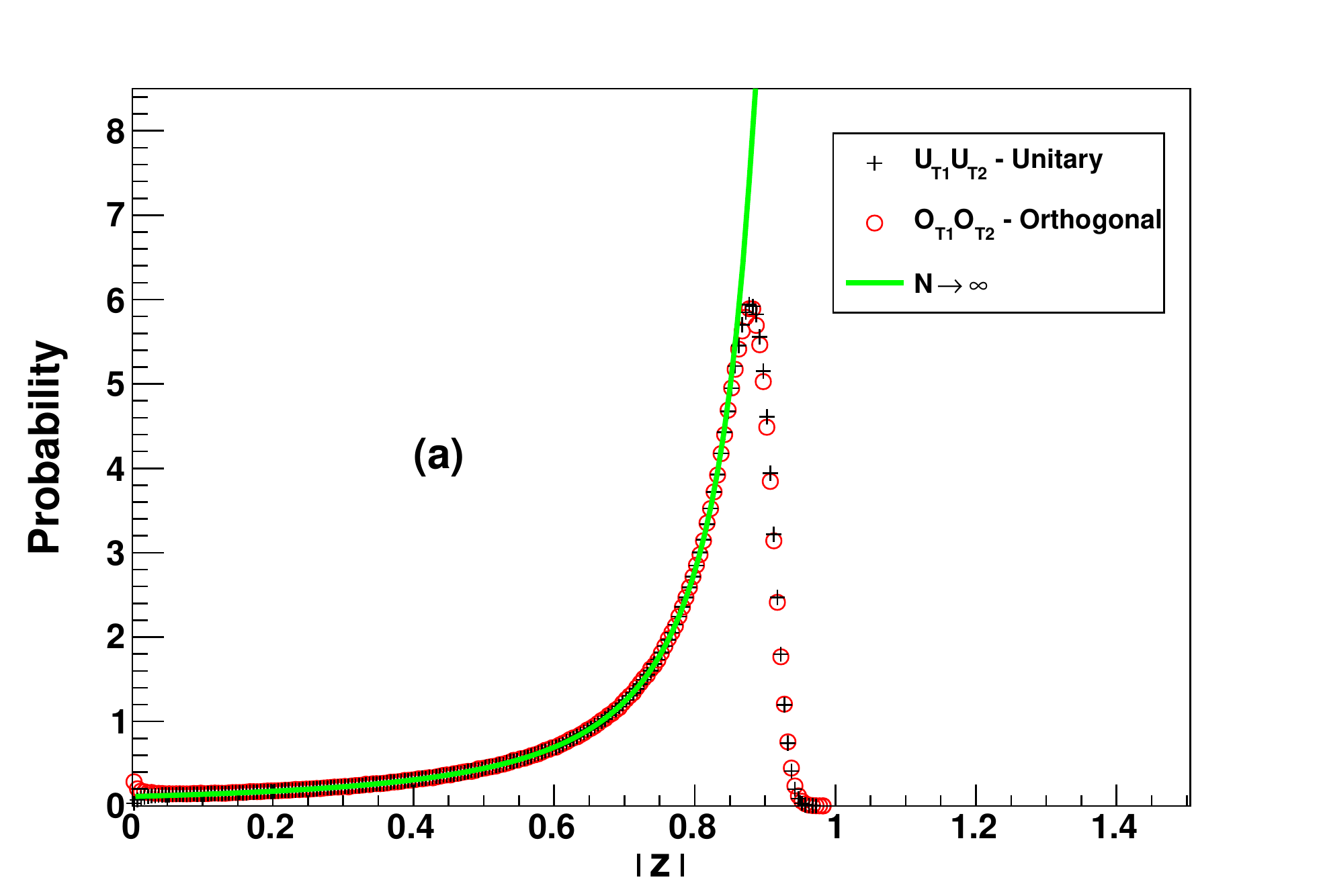} 
\includegraphics[width=59.1mm]{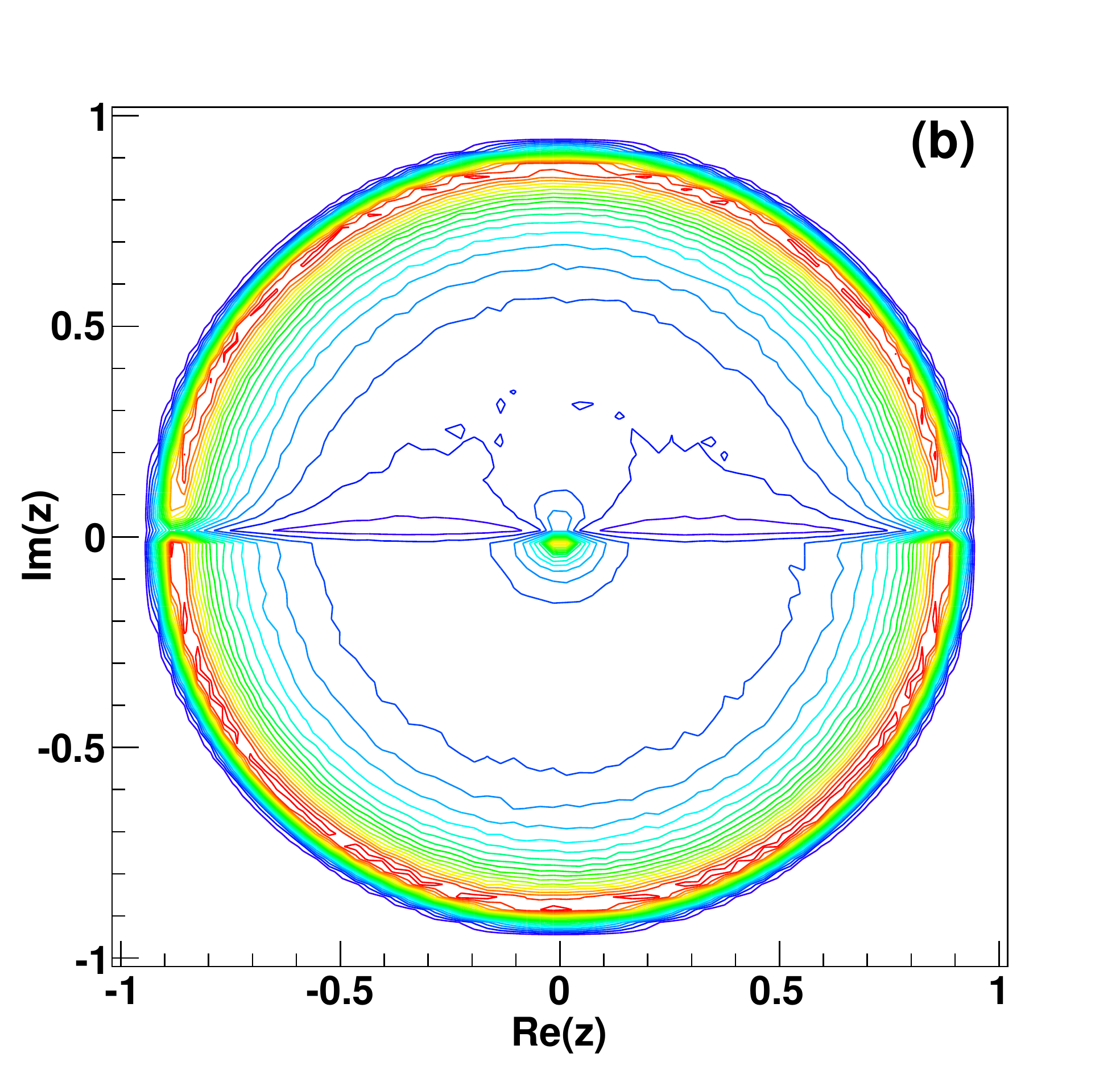} 
\includegraphics[width=59.1mm]{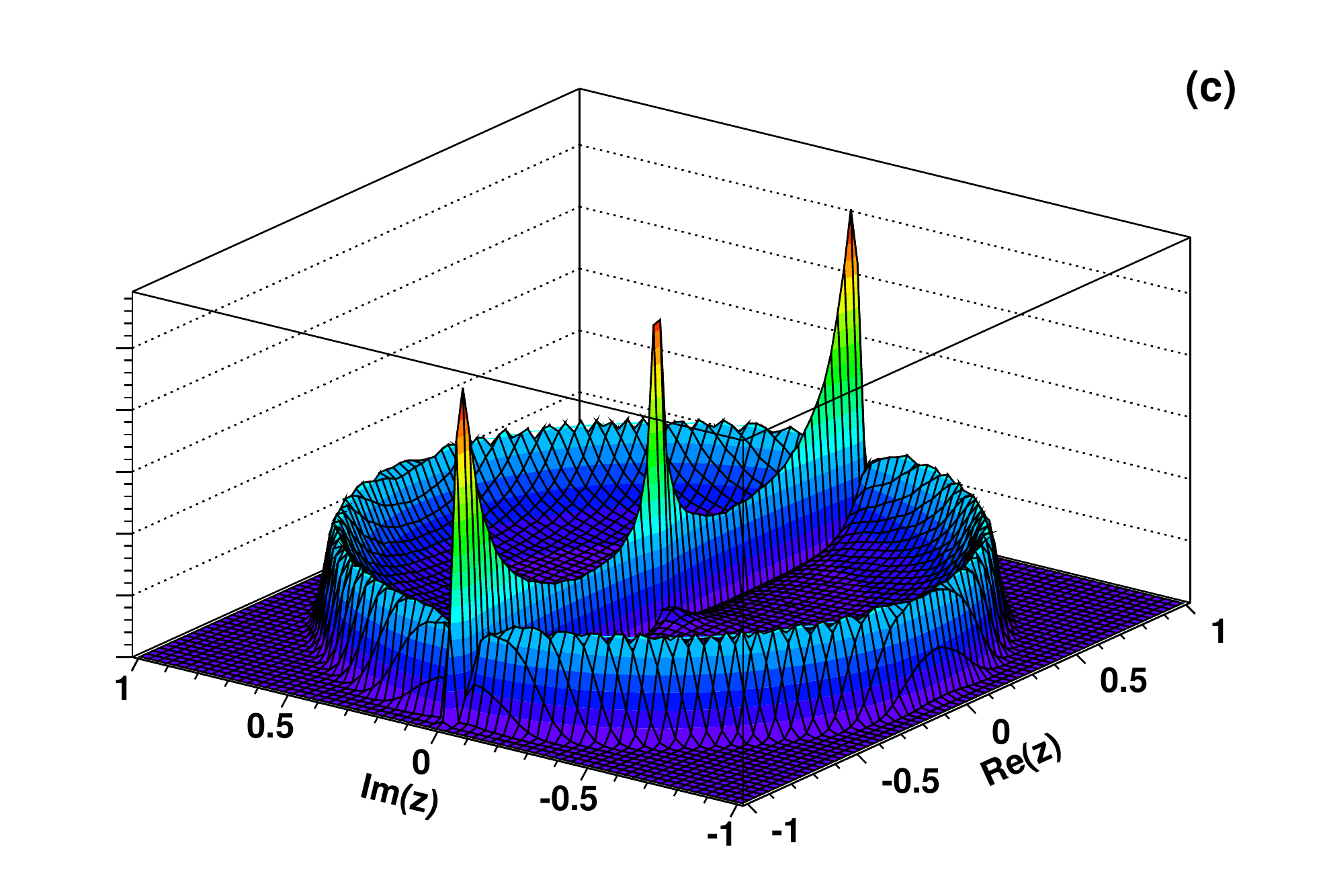}
\caption{(Color online) Numerical comparison of the eigenvalues of product of two truncated orthogonal and unitary matrices. (a) The radial part $p(r)=F'(r)$ (\ref{density}) of the mean spectral density $\rho\left(z,\bar{z}\right)$ for unitary (black crosses) and orthogonal (red circles) matrices for $N=100$, $\kappa=\frac{1}{9}$. Each histogram is made for $9\times10^{6}$ eigenvalues. 
The theoretical prediction for $N\rightarrow\infty$ is shown for comparison (solid green line). (b) Full eigenvalue distribution of orthogonal (upper half of complex plane) and unitary (lower half of complex plane) truncated matrices for $N=100$, $\kappa=\frac{1}{9}$. (c) Full eigenvalue distribution of orthogonal truncated matrices for same $N$,$\kappa$ parameters. The real eigenvalue band is clearly visible.}
\label{fig:numorth}
\end{figure}

We also performed numerical simulations for the products of truncated orthogonal matrices
as an example of multiplication of IOE matrices. In the
large $N$ limit both the IUE and IOE densities are expected to 
have the same limiting distribution while for finite $N$ the distribution in
the IOE case is expected to display a characteristic pattern that
weakly breaks the circular symmetry of the eigenvalue distribution on the complex 
plane. More precisely, one expects that a fraction of eigenvalues accumulates 
on the real axis and disappears from a narrow depletion region close to the axis. 
The effect was first discussed for real Girko-Ginibre matrices \cite{e} 
and later also for orthogonal truncated matrices \cite{ksz}. It is known to be
a finite size effect in the sense that the fraction of eigenvalues 
forming the pattern tends to zero for $N\rightarrow \infty$
so the full circular symmetry of the eigenvalue density is restored in the limit.
In fact, this is exactly what we see in our numerical simulations of the product
of truncated matrices. First we observe that the radial distribution of eigenvalues 
of product of two truncated unitary matrices is identical to the case of truncated orthogonal matrices except in a small region close to $r=0$ (see fig. \ref{fig:numorth}.a).
In figure \ref{fig:numorth}.b we compare finite size distributions for
the product of IUE (lower part) and IOE (upper part). We see that the IUE distribution is circularly symmetric up to the statistical noise 
while the IOE distribution has an elongated shape close to the real axis,
as expected. Finally in fig. \ref{fig:numorth}.c we show the full spectrum
on which one can clearly see an accumulation of eigenvalues on the real axis.

\section*{Discussion}
In this note we have shown a simple, (and as far as we know) new relation between the spectral properties of the product of  $n$ identically distributed isotropic random matrices from the given IUE ensemble and spectral properties of $n$-th power of a single matrix from this ensemble. 
We stress a nonintuitive aspect of this result that tells us that 
independent matrices, when multiplied, give the same eigenvalue 
density as the product of fully correlated (identical) matrices.
In a sense it is a self-averaging effect: a single random matrix from
a isotropic ensemble is a good representative  
to describe products
of matrices from this ensemble in the limit $N\rightarrow \infty$. -

We have supplemented our analytic proof with extensive numerical simulations, allowing us to see how the finite size effects vanish in the thermodynamical limit. For Girko-Ginibre finite size effects agree with those conjectured in \cite{b1,b2}. 

Our result elucidates the transparent  analytic structure noted  in
several recently published papers on the products of random
matrices~\cite{bjw,agt1,agt2,b1,b2,bjn,os,sz,bmj,bbcc,ms,gv,fk,djl}
and provides a powerful tool for the  derivation of  similar results for
products of some application-designed, isotropic random matrices of large
(infinite) size.

\subsection*{Acknowledgments}
We thank Yan Fyodorov and Karol Zyczkowski 
for stimulating discussions. This work was supported by 
the Polish Ministry of Science Grant No. N N202 229137 (2009-2012)
and by the Grant DEC-2011/02/A/ST1/00119 of National Centre of Science.

\section*{Appendix A}
In this appendix we briefly recall basic facts about the S-transform, 
introduced by Voiculescu in free random probability \cite{v}.
Consider a Hermitian random matrix $a$. One usually
defines the Green function
\be
G_a(z) = \lim_{N\rightarrow \infty} 
\frac{1}{N} \left\langle \mbox{Tr} (z-a)^{-1} \right\rangle =
\int d\lambda \frac{\rho_{a}(\lambda)}{z-\lambda}  \ .
\ee
that is directly related to the eigenvalue density $\rho_a(\lambda)$. 
Note that the density is a function of a real variable while the 
Green's function is a function of a complex variable.
The Green's  function generates moments $\mu_{ak}$ (if they exist) 
\be
G_a(z) = \frac{1}{z} + \sum_{k=1}^\infty \frac{\mu_{ak}}{z^{k+1}}
\ee
of the eigenvalue density
\be
\mu_{ak} = \lim_{N\rightarrow \infty} 
\frac{1}{N} \left\langle \mbox{Tr} \; a^k \right\rangle = 
\int d\lambda \rho_a(\lambda) \lambda^{k} \ .
\ee
Sometimes it is more convenient to use another generating function,
given by a power series in $z$ rather than in $1/z$:
\be
\psi_a(z) = \frac1z G_a\left(\frac1z\right) - 1 = \sum_{k=1}^\infty \mu_{ak} z^{k}
\label{psi_def}
\ee
and to introduce its functional inverse $\chi_a$:
\be
\chi_a(\phi_a(z))=\phi_a(\chi_a(z))=z
\label{chi_def}
\ee
which can also be expressed as a power series in $z$ if the first
moment is nonzero: $\mu_{a1} \ne 0$. The S-transform for the matrix $a$
is related to the $\chi$-transform as 
\be
S_{a}(z) = \frac{1+z}{z} \chi_a(z) \ .
\label{S_def}
\ee
The relevance of the S-transform in free probability is related
to the fact that it allows one to concisely formulate the law of 
free multiplication. The S-transform of the product of two free
(independent) matrices from invariant ensembles is a product of the 
S-transforms of individual matrices:
\be
\label{SS}
S_{ab}(z) = S_a(z) S_b(z)
\ee
The multiplication law was formulated in free random probability \cite{v}
but it also can be rederived in random matrix set-up 
using field theoretical techniques for the summation of 
planar Feynman diagrams and it can be generalized to 
non-hermitian matrices \cite{bjn}.

\section*{Appendix B}

In this appendix we rederive 
the distribution of a single unitary truncated matrix (\ref{single_trunc})
using free probability and the Haagerup-Larsen theorem. 
We first construct the density of an $(N+L)\times (N+L)$ matrix $y=p u$ where 
\be
p = \mbox{diag}(\underbrace{1,\ldots,1}_N,\underbrace{0,\ldots,0}_L)
\ee
is a projection matrix and $u$ is 
an Haar unitary matrix of dimensions $(N+L)\times (N+L)$.
In order to calculate the S-transform for the projector $p=p^2$
we first observe that all moments of $p$ are equal $\mu_k = N/(N+L)$.
Hence $\psi_p(z) = \frac{N}{N+L} \frac{z}{1-z}$ (\ref{psi_def}), 
$\chi_p=\frac{z}{N/(N+L)+z}$ (\ref{chi_def}) and eventually
(\ref{S_def})
\be
S_p = \frac{1+z}{N/(N+L)+z} \ .
\ee
Inserting this to (\ref{HL1}) we find
\be
F_y(r) = \frac{L}{N+L} \frac{1}{1-r^2}  \qquad \mbox{for} \quad  r\le \sqrt{N/(N+L)}
\ee
and $1$ otherwise. We see that $F_y(0)=L/(L+N)$. This means that there 
are $L$ eigenvalues equal zero. They are inherited from the zero 
eigenvalues of the projector. We can now reduce dimensionality
of the matrix $y$ by removing $L$ zero eigenvectors. 
The remaining matrix $x$ that has no trivial 
zero eigenvalues. This gives the result given in equation (\ref{single_trunc})  
for a single truncated matrix
\be
F_x(r) = \frac{N+L}{N}\left( F_x(r) - \frac{L}{N+L}\right) = 
\frac{L}{N} \frac{r^2}{1-r^2}  \qquad \mbox{for} \quad  r\le \sqrt{N/(N+L)}
\ee
and one otherwise. The term $-L/(N+L)$ removes $L$ zero eigenvalues out of $(N+L)$ eigenvalues, and the factor $(N+L)/N$ restores the total normalization.

\end{document}